\documentstyle[12pt]{article}

\def \be{\begin{equation}}
\def \ee{\end{equation}}
\def \bea{\begin{eqnarray}}
\def \eea{\end{eqnarray}}
\begin{document}
{\hskip 9cm  }      
{\hskip 1cm  }
\\
\begin{center}
{\Large{\bf{ Turbulence With Pressure
}}}
\vskip .5cm   
{\large{A. R. Rastegar${}^1$, M. R. Rahimi Tabar${}^{2,3}$, P. Hawaii${}^1$}}
\vskip .1cm
{\it{1) Department of Physics, Tabriz University, Tabriz, Iran
\\2) Dept. of Physics , Iran  University of Science and Technology,\\
Narmak, Tehran 16844, Iran.
\\3) Institue for Studies in Theoretical Physics and 
Mathematics
\\ Tehran P.O.Box: 19395-5746, Iran.}}
\end{center}
\vskip .5cm
\begin{abstract}

 We investigate the exact results of the Navier-Stokes equations using the 
methods
developed by Polyakov. It is shown that when the velocity field and the 
density are not independent, the Burgers equation is obtained leading to 
 exact 
N-point generating functions of velocity field. Our results show that,  
the operator product expansion has to be generelized 
both in the absence and the presence of pressure. We find a method to 
determine the extra terms in the operator product expansion and derive its
coefficients and find the first correction to probablity distribuation function. 
In the general case and for small pressure, we solve the problem 
perturbatively and find the probablity distribuation function
for the Navier-Stokes equation
in the mean field approximation.  

\end{abstract} 
\vskip .5cm

PACS number 47.27.AK
\newpage
{\bf 1- Introduction}

A theoretical understanding of turbulence has eluded physicists for a 
long time. A statistical theory of turbulence has been put
forward by Kolmogorov [1], and further developed by others [2-4].
This is to model turbulence using stochastic
partial differential equations.  In this direction, Polyakov [5] has 
recently offered a field theoretic method for deriving the 
probability distribution or density of states in (1+1)-dimensional 
in the problem of randomaly driven Burgers equation. 
The importance of the Burgers equation is that, it is the simplest
equation that resembles the analytic structure of the N-S equation, at least formally, 
within the scope of applicability of the Kolmogorov`s arguments.

Polyakov 
formulates a new method for analyzing the inertial range correlation 
functions based on the two important ingredients in field theory and 
statistical physics namely the operator product expansion (OPE) and 
anomalies.
He argues that in the limit of high Reynold's number because of 
existence of singularities at coinciding point, dissipation remains 
finite and 
all sublleading terms give vanishing contributions in the inertial range.
By using the OPE one finds the leading singularities and
can show that this approach is self-consistent.

Here we generlize Polyakov`s approach to the Burgers turbulence [5] 
in the presence of the pressure (i.e. the Navier-Stokes equation). 
We consider two distinct situations, firstly when the velocity field and  
the density are not independent, and secondly when the gradient of the 
pressure is small. In the first case we show that by a change of variables
one can transform this problem to the Burgers equation and fllowing [6]
we find the exact N-point generating function of the velocities.
For the second case we calculate the effect of pressure perturbatively and 
find the assymptotic behavior of probablity disturbuation function 
in the first approximation.

On the other hand, using the results of [6], we calculate the OPE coefficiente  
proposed by Polyakov, and show that we have to correct the OPE both in the 
presence and the absence of the pressure.
We write the generlized OPE and find the first correction to the PDF which 
was initially  found by Polyakov. 

The paper is organised as follows. In section two we show 
that when the velocity and density are not independent the governing
equation reduces to the Burgers equation. In section three we find 
the N-point generating functions for the velocity field and show that 
we have to generlize the OPE proposed by Polyakov. We find the OPE 
coefficients and derive the first correction to probablity disturbuation 
function.  In section four we consider the effect of 
pressur perturbatively and find the probablity disturbuation 
function in the mean field approximation.
\\
{ \bf 2-  The Compressible fluid and the Burgers equation} 

We begin our discussion with the system describing compressible flows in an 
ideal (polytropic) gas ignoring dissipative effect [7], viz.
\be 
\rho_t + v \rho_x +\rho v_x =0
\ee
\be
\rho(v_t + v v_x) + p_x =0
\ee
\be
p=k \rho^{\gamma}, \hskip 1cm s=const.
\ee
Here $\rho$, $v$ and $p$ are the density, velocity field and pressure, 
respectively. The $s$ is entropy, assumed to be constant, $\gamma=\frac{C_p}{C_v}$,
the ratio of specific heats and $k$ is a constant.

Also in inertial range, we ignore the driving force, 
which pumps energy in the large scale of system.
The system of eqs.(1-3) has essentialy two variables ($\rho$ and $v$), it
is nonlinear and difficult to handel in complete generality. Nevertheless 
there has been considerable analytic interest in this system. 
In particular, some progress can be made by seeking simple wave solutions 
where one of the dependent variables is a functions of the others.
Rewriting eq.(1-3) in terms of $v$ and $\rho$ only by introducing
the square 
of the speed of sound, $a^2=(\frac {\partial p}{\partial \rho})_{s=s_0} = 
k \gamma \rho^{\gamma -1}$, we have: 

\be
\rho_t + v \rho_x +\rho v_x=0
\ee
\be
v_t + v v_x +\frac{a^2(\rho)}{\rho} \rho_x =0
\ee

Now, we assume that $v=v(\rho)$, so that eqs.(4) and (5) changes to the 
following equations:
\be
\rho_t + (v+ \rho v \prime ) \rho_x =0
\ee
\be
\rho_t + (v + \frac{a^2}{\rho v \prime})\rho_x=0
\ee
Here the prime denotes differentiation with respect to $\rho$.
This system of linear algebraic equations in $\rho_t$ and $\rho_x$ has a 
non-trivial solution provided the determinant of the coefficient matrix 
vanishes so that:
\be
v\prime = \pm \frac{a}{\rho} = \pm a_0 (\frac{\rho}{\rho_0})^{\frac{\gamma -1}{2}}
\frac{1}{\rho}
\ee
The system (6-7), then reduces to one of the equations:
\be
\rho_t + (v \pm a) \rho_x = 0
\ee
where
\be
v(\rho)=\int_{\rho_0} ^{\rho} \frac{a(\rho)}{\rho} d \rho =\frac{2}{\gamma -1} 
\{a(\rho)-a_0\} 
\ee
We restrict our attention to waves moving to the right thus choosing 
the plus sign in eqs.(8) and (9).
The corresponding equation for $v$ follows easily from multiplication of eq.(6)
by $v\prime (\rho)$ and writing the result in terms of $v$ via eq.(10). Therefore
we obtain:
\be
v_t + (a_0 + \frac{\gamma +1}{2} v ) v_x =0
\ee
Cole generlizes the eq.(11) in the presence of small kinematical viscosity 
and have found
the following equation for $v$ [8]:
\be
v_t + (a_0 + \frac{\gamma +1}{2} v ) v_x = \nu v_{xx}
\ee

This equation can be transformed into the Burgers equation by simple
change of variables.
We define $u= a_0 + \frac{\gamma +1}{2} v$ and in the presence of 
some driving force in large scale eq.(12) transfomed to:
\be
u_t + uu_x = \nu u_{xx} + \bar f(x,t)
\ee
where  $\bar f(x,t)$ 
is a 
Gaussian random force with the following correlation:
\be
<\bar f(x,t) \bar f(x^{'},t^{'})> = (\frac{\gamma+1}{2})^2 k (x-x^{'}) \delta (t-t^{'}) 
\ee
The transformation, $u(x,t) = -\lambda \partial_x h(x,t)$ maps eq.(13) to the 
well knwon Kardar-Parisi-Zhang equation [9],
\be
\partial_t h = \nu \partial_{xx} h + {\lambda \over 2}[\partial_x h]^2 +
\bar f_2(x,t)
\ee
which gives the general features of many complex system [9,10].
It is noted that the parameter $\lambda$ that appears in the above 
transformation is not 
renormalized under any renormalization procedure [11].\\
\newpage
{\bf 3-  The Exact N-Point Generating Functions and the Generelized OPE} 

To statistical description of eq.(13) following 
 Polyakov [5] consider the following generating functional
\be 
Z_N(\lambda_1, \lambda_2,\ldots \lambda_N, x_1,\ldots x_N) = <\exp (
\sum^N_{j=1} \lambda_j u(x_j,t))>
\ee
Noting that the random force $\bar f(x,t)$ has a Gaussian distribution, 
$Z_N$ satisfies a closed differential 
equation provided that the viscosity  $\nu$ tends to zero:
\be
\dot{Z_N} + \sum \lambda_j {\partial \over {\partial \lambda_j}}({1\over 
{\lambda_j}}{\partial Z_N \over{\partial x_j}}) = \sum \bar k(x_i-x_j) \lambda_i
\lambda_j Z_N +D_N
\ee
where $D_N$ is :
\be
D_N = \nu \sum \lambda_j <u^{''}(x_j,t) \exp \sum \lambda_k u(x_k,t)> 
\ee
and 
\be
\bar k(x_i-x_j)=(\frac {\gamma+1}{2})^2 k(x_i-x_j)
\ee

To remain in the inertial range we must, however, keep
$\nu$ infinitesimal but non-zero. Polyakov argues that the anomaly mechanism 
implies that infinitesimal viscosity produces a finite effect. To compute 
this effect Polyakov makes the F-conjecture, which is the existence of 
an operator 
product expansion or the fusion rules. The fusion rule is the statement 
concerning the behaviour of correlation functions, when some subset of 
points are put close together.

Let us use the following notation;
\be
Z(\lambda_1, \lambda_2,\ldots , x_1,\ldots x_N)= <e_{\lambda_1}(x_1)\ldots
e_{\lambda_N}(x_N)>
\ee
then Polyakov`s F-conjecture is that in this case the OPE has the 
following form,
\be
e_{\lambda_1}(x+y/2) e_{\lambda_2}(x-y/2) = A(\lambda_1, \lambda_2, y)
e_{\lambda_1+\lambda_2}(x)+B(\lambda_1, \lambda_2, y){\partial \over \partial x}
e_{\lambda_1+\lambda_2}+o(y^2)
\ee

This implies that $Z_N$ fuses 
into functions $Z_{N-1}$ as we fuse a couple of points together. The 
F-conjecture
allows us to evaluate the following anomaly operator (i.e. the 
$D_N$-term in eq.(5)),
\be
a_\lambda(x)= \lim _{\nu\rightarrow 0}\nu (\lambda u^{''}(x) \exp 
(\lambda u(x))
\ee
which can be written as:
\be
a_\lambda(x)= \lim _{\xi,y,\nu\rightarrow 0}\lambda \nu {\partial^3\over 
{\partial \xi \partial y^2}} e_\xi (x+y) e_\lambda(x)
\ee
As discussed in [3] the possible Galilean invarant expression is:
\be
a_\lambda(x) = a(\lambda) e_\lambda(x) + \tilde{\beta}(\lambda){\partial
\over \partial x}e_\lambda(x)
\ee
Therefore in steady state the master equation takes the following form,
\bea
\sum ({\partial \over \partial \lambda_j} - \beta(\lambda_j)){\partial \over 
\partial x_j} Z_N -\sum \bar k(x_i -x_j) \lambda_i \lambda_j Z_N &=& \sum 
a(\lambda_j) Z_N \cr
\beta(\lambda) &=& \tilde{\beta}(\lambda)+{1 \over \lambda}
\eea

Let us consider following correlation for $\bar f$ in $k$-space as follows [12]:
\be
<\bar f(k,t) \bar f(k^{'},t^{'})> =\frac {L}{2\pi} \bar k_0  (\frac{\gamma+1}{2})^2\delta 
(k^2 - \frac{1}{L^2}) 
\delta (k+k^{'}) \delta (t-t^{'})        
\ee
therefore we obtain:
\be
<\bar f(x,t) \bar f(x^{'},t^{'})> = \bar k_0 (\frac{\gamma+1}{2})^2 
cos(\frac { x-x^{'}}{L}) \delta (t-t^{'})        
\ee
In the inertial range where $ x_i -x_j << L$, we can expand the 
r.h.s. of eq.(27) and find
\be
\bar K(x_i - x_j) =  \bar k_0  (\frac{\gamma+1}{2})^2 
(1- \frac{(x_i - x_j)^2}{2 L^2} ) 
\ee

Therefore we find the following explicit form of $Z_2$ for 
$\bar k(x_i -x_j)$ given by eq.(28): 
\be
 Z_2(\mu y) = e^{\frac {\gamma +1}{3}(\mu y)^{3/2}}
\ee
and the following experssion for density of states as the Laplace transform
of $Z_2$ ;
\be                         
W(u,y) = \int_{c-i\infty}^{c+i\infty} {d\mu \over 2\pi i} e^{-\mu u} 
Z_2(\mu y)
\ee
where  $$\mu=2 (\lambda_1 -\lambda_2), \hskip .5 cm y=x_1-x_2.$$
It can be easly shown that with the following definition  of 
variables  Polyakov`s
master equation (i.e. eq.(25) with the scaling conjecture[5] is:
\be
\{{\partial^2 \over {\partial \mu_2 \partial y_2}}+ {\partial^2 \over 
{\partial \mu_3 \partial y_3}}-(\frac{\gamma+1}{2})^2 (y_2 \mu_2 +y_3 \mu_3)^2\} f_3 =0
\ee
where
$$f_3 = (\lambda_1 \lambda_2 \lambda_3)^{-b_3}Z_3$$
\bea
y_1 &=& {x_1+x_2+x_3 \over 3} \hskip 0.5cm y_2= x_1-{x_2+x_3 \over 2} 
\hskip 0.5cm y_3 =x_2-x_3 \cr
\mu_1 &=& {\lambda_1+\lambda_2+\lambda_3 \over 3} \hskip 0.5cm 
\mu_2= {3 \over 2}(\lambda_1-{\lambda_2+\lambda_3 \over 2})
\hskip 0.5cm \mu_3 =2(\lambda_2-\lambda_3)
\eea
Now we set $f_3$ as;
\be
f_3 = \mu_2^{S_2} \mu_3^{S_3}g_3(\mu_2 y_2,\mu_3 y_3)
\ee
inserting in eq.(15) results in
\be
g_3(\mu_2 y_2, \mu_3, y_3) = e^{\frac{\gamma + 1}{3}(\mu_2 y_2+ \mu_3 y_3)^{3/2}}
\ee
and $$ S_2 = S_3 =-5/4$$
Now if we use the following transformation;
\bea
y_1&=&{{x_1 +x_2 +x_3 +\ldots x_N}\over N}\cr
y_2&=&x_1 -{{x_2 +x_3 +\ldots x_N}\over N-1}\cr
y_3&=&x_2 -{{x_3 +x_4 +\ldots x_N}\over N-2}\cr
and \hskip 1cm y_N&=& x_{N-1}-x_N 
\eea
and
\bea
\mu_1 &=&{{\lambda_1 + \lambda_2 +\ldots +\lambda_N}\over N}\cr
\mu_2 &=&{N\over N-1}[\lambda_1-{{\lambda_2 + \lambda_3 +\ldots +\lambda_N}
\over {N-1}}]\cr
\mu_3 &=&{N-1 \over{N-2}}[\lambda_2-{{\lambda_3 + \lambda_4 +\ldots 
+\lambda_N
}\over {N-2}}]\cr
and \hskip 1cm \mu_N&=& 2(\lambda_{N-1} - \lambda_N)
\eea
we get the following partial differential equation for $f_N$;
\be
\{{\partial^2 \over {\partial y_2 \partial \mu_2}} + \ldots +{\partial^2
\over {\partial y_N \partial \mu_N}}\} f_N - (\frac{\gamma+1}{2})^2 
(y_2 \mu_2+\ldots +y_N \mu_N)^2
f_N = 0
\ee
which is solved by:
\be
f_N = (\mu_2 \mu_3 \ldots \mu_N)^{-[{2N-1 \over {2(N-1)}}]} 
\exp^{\frac {\gamma +1}{3}(
\mu_2 y_2 + \ldots + \mu_N y_N)^{3/2}}                                          
\ee

The parameter $b_n$ for the $N$-point generating functions can be determined
by the requirment that $Z_N$ must be finite in the limit $\lambda_i \rightarrow 0$.  
Therefore we find:
\be
b_N = \frac {2N-1}{2N}
\ee
It follows that the behavior of N-point correlation functions of $v$ is: 
\be
 G^{(N)} (x_1, \cdots ,x_N)\sim lim_{\lambda \rightarrow 0} \lambda^{-N} 
 \sum_{k=0} ^N a_k ^{(N)}   (\lambda x)^{\frac {3k}{2}} 
\ee
where $a_k^{(N)}$ are some constants.
\\

Now we try to find the OPE coefficients in eq.(21).

This can be done by using the $Z_3$ and put  $x_2$ and $x_3$ close together.
Here we calculate the OPE coefficiens for the case $\gamma=1$ i.e. the 
ordinary Burgers turbulence and for arbitrary $\gamma$ we need only small 
modifiation. $Z_3$ and $Z_2$ are given by eqs.(32) and (29).
In eq.(32) we take $x_3 = x_2 - 2\epsilon$ and it is easy to show that in the limit
$\epsilon \rightarrow 0$ we find:
\be
Z_3 \Rightarrow (\lambda_1 \lambda_2 \lambda_3)^{5/6}(\mu_2 \mu_3)^{-5/4}
e^{2/3(\mu_2 y_2+ \mu_3 y_3)^{3/2} 
(1-\frac {1}{8} \frac {2 \lambda_1 - 5 \lambda_2 - 5  \lambda_3}
{ \lambda_1 -  \lambda_2 -  \lambda_3} + \frac {1}{8} \frac {6 \lambda_1 + 
13 \lambda_2 - 19  \lambda_3}
{( \lambda_1 -  \lambda_2 -  \lambda_3)(x_1-x_2)} \epsilon )^{3/2}} 
\ee
where $\lambda_3= - (\lambda_1 + \lambda_2)$. Now we can 
expand the exponent of eq.(41) in termes of $\epsilon$, therefore we find:
\bea
Z_3 &=& (\lambda_1 \lambda_2 \lambda_3)^{5/6}(\mu_2 \mu_3)^{-5/4}
e^{2/3(\mu_2 y_2+ \mu_3 y_3)^{3/2}} (1- \frac{1}{4} (2 \lambda_1
-5 \lambda_2 - 5 \lambda_3 )\nonumber \\&(&2(\lambda_1-\lambda_2-\lambda_3))^{1/2} 
(x_1 - x_2)^{3/2}
\frac{1}{4} (6\lambda_1 + 13 \lambda_2 - 19 \lambda_3 )\nonumber \\&(&2(\lambda_1                                            
-\lambda_2 -\lambda_3))^{1/2} (x_1 - x_2)^{1/2} \epsilon  +  O(\epsilon)^2 )
\eea

Comparing with eq.(21) we find that:
\be
A(\lambda_1,\lambda_2,\epsilon) = (-\lambda_1 \lambda_2 (\lambda_1 + \lambda_2))^{5/6}
(\frac{9}{2} \lambda_1 (\lambda_1 + 2 \lambda_2))^{- \frac{5}{6}}
\ee
\be
B(\lambda_1, \lambda_2, \epsilon)=- \frac{A(\lambda_1,\lambda_2,\epsilon)}{16} 
(\frac{25\lambda_1 + 32 \lambda2}{\lambda_1}) \epsilon
\ee
\be
C(\lambda_1 , \lambda_2 , \epsilon)= \frac{7}{8} A(\lambda_1,\lambda_2,\epsilon)
\ee
Where $C(\lambda_1 , \lambda_2 , \epsilon)$  can treated as the third terms
in the OPE of eq.(21) and are given in the following 
modified OPE:
\bea
e_{\lambda_1}(x+ \epsilon /2) e_{\lambda_2}(x- \epsilon /2)& =& A(\lambda_1, \lambda_2, \epsilon)
e_{\lambda_1+\lambda_2}(x)+B(\lambda_1, \lambda_2, \epsilon){\partial \over \partial x}
e_{\lambda_1+\lambda_2} \nonumber \\ &+& C(\lambda_1 , \lambda_2 , \epsilon)\frac {\partial 
}{\partial \lambda} e_{\lambda_1+\lambda_2} + O(\epsilon ^2) 
\eea

Now one can show that the modified OPE leads to following equations for 
$a_\lambda (x)$ and $Z_2$:
\be
a_\lambda(x) = a(\lambda) e_\lambda(x) + \tilde{\beta}(\lambda){\partial
\over \partial x}e_\lambda(x) + \tilde{\gamma}(\lambda){\partial
\over \partial \lambda}e_\lambda(x) +  \cdots
\ee
\be
(\partial_{\mu} - \frac {2b}{\mu})\partial_y Z_2 + c \mu \partial_\mu Z_2 -
\mu^2 y^2 Z_2=0
\ee
where we have used $\tilde{\beta}(\lambda)= \frac{b-1}{\lambda}$ , $a=0$ according
[4], and scaling arguments show that $ \tilde{\gamma}(\lambda)$ has 
the following form:
\be
\tilde{\gamma}(\lambda)= c \lambda
\ee
The eq.(49) have the following assymptotic behavior:
\be
 Z_2(\mu y) = e^{\frac {\gamma +1}{3}(\mu y)^{3/2} - \frac{c}{2} (\mu y)}
\ee
It appers that in the limit $c\rightarrow 0$, we find Polyakov`s
result for $Z_2$. Indeed our results shows that we can find the exact 
$Z_2$ for the Burgers equation by calculating the exact form of OPE,
proposed by Polyakov by methods disscussed above.
\\

{\bf 4-  Perturbative Calculations of the Pressure}  

To consider the general situation we have the following set of 
equations describing the compressible fluid:

\be
\rho_t + v \partial_x \rho +\rho \partial_x v = 0
\ee
\be
v_t + v \partial_x v + \frac {\partial_x p}{\rho} = \nu \partial_x ^2 v + f(x,t)
\ee  
\be
p= k \rho^{\gamma}
\ee
By using the eq.(53) we find:
\be
\rho_t + v \partial_x \rho +\rho \partial_x v = 0
\ee
\be
v_t + v \partial_x v + k \gamma \rho^{\gamma - 2} \partial_x \rho =               
\nu \partial_x ^2 v + f(x,t)
\ee
where the $\nu$ is the viscosity.
We rewite the eqs.(54) and (55) for different points $x_1$ and $x_2$ 
and multiplying the resulting equations by $\rho(x_2) , \lambda_1 \rho(x_2),
\rho(x_1) $ and  $ \lambda_2 \rho(x_1)$ respectively, and all of
them to 
\be
e^{\lambda_1 v(x_1 ,t) + \lambda_2 v(x_2 ,t) }
\ee
and averaging with respect to the external force we find:
\bea
\sum_{j=1} ^2 ({\partial \over \partial \lambda_j} - \beta(\lambda_j)){\partial \over 
\partial x_j} Z_2 -\sum_{j=1} ^2 \bar k(x_i -x_j) \lambda_i \lambda_j Z_2 + F_2 
&=& \sum_{j=1} ^2 
a(\lambda_j) Z_2 + \Lambda_2 Z_2 \cr
\beta(\lambda) &=& \tilde{\beta}(\lambda)+{1 \over \lambda}   
\eea
where
\be
Z(\lambda_1, \lambda_2, x_1,x_2)= < \rho (x_1) \rho (x_2) e^{\lambda_1 v(x_1) + 
\lambda_2 v(x_2)}>
\ee
\be
F_2=<k \gamma \rho (x_1) \rho (x_2) e^{\lambda_1 v(x_1) + \lambda_2 v(x_2)}
[\lambda_1 \rho^{\gamma-2}(x_1) \partial_{x_1 } \rho(x_1) + 
\lambda_2 \rho^{\gamma-2}(x_2) \partial_{x_2} \rho(x_2)]>          
\ee
and  
\be
\Lambda_2 = mean (\sum_{i=1} ^2  \tilde{\beta}(\lambda_i) \frac {1}{\rho(x_i)}
\partial_{x_i} \rho(x_i) )
\ee
For small pressure we can approximate the eq.(59) in the mean field 
approximation as: 
\be
F=-\eta Z, \hskip 2cm  \eta=mean (-k \gamma 
[\lambda_1 \rho^{\gamma-2}(x_1) \partial_{x_1 } \rho(x_1) + 
\lambda_2 \rho^{\gamma-2}(x_2) \partial_{x_2} \rho(x_2)] )         
\ee

By using $\beta(\lambda)=\frac{b}{\lambda}$ and $a(\lambda)=0$ we find the following
asmyptotic expression for $Z_2$:
\be
Z_2=e^{\frac {2}{3} (\mu y)^{3/2} - (\Lambda_2 + \eta) (\mu y )^{-1/2}} \hskip 1cm
\mu y \rightarrow \infty 
\ee
\\
{\bf Acknowledgements:} We would like to thank M. A. Jafarizadeh, M. Khorrami and 
S. Rouhani for valuable 
discussions. 

\end{document}